# Experimental evidence of a crossover between cooperative relaxation and liquid growth dynamics

Ana Vila-Costa, Marta Gonzalez-Silveira[$], Cristian Rodríguez-Tinoco, Marta Rodríguez-López, Javier Rodríguez-Viejo[$]

*Departament de Física. Facultat de Ciències, Universitat Autònoma de Barcelona, 08193, Bellaterra, Spain*

*Catalan Institute of Nanoscience and Nanotechnology (ICN2), CSIC and BIST, Campus UAB, Bellaterra, 08193, Barcelona, Spain*

**Abstract:**

In stark contrast with the conventional understanding of the glass transition, where the transition from glass to liquid appears as a dynamic process where atoms/molecules cooperatively relax into the equilibrium phase, we experimentally show that the nature of the glass transition depends at a given temperature on the ratio between the relaxation time of the glass, $\tau_{glass}$, taken as its transformation time, and the alpha relaxation time, $\tau_\alpha$. Although the relaxation of liquid-cooled glasses is not totally synchronous, due to the existence of a distribution of relaxation times, there has been no clear observation of phase separation. However, at temperatures at which $\tau_{glass}/\tau_\alpha$ is large, high mobility regions nucleate into the liquid phase that subsequently grow by dynamic facilitation before – or while - cooperative glass relaxation sets into play. On the contrary, at temperatures associated to smaller $\tau_{glass}/\tau_\alpha$ the glass transition proceeds by cooperative relaxation dynamics all-across the material. This behavior is independent of the experimental procedure or protocol to produce the glass.

[$]**Corresponding authors:** marta.gonzalez@uab.cat**,** javier.rodriguez@uab.es

Understanding the physics of glass formation upon cooling a liquid and its converse effect, the transition to a supercooled liquid upon heating the glass, is still a challenge despite the intense experimental and theoretical research of the last 100 years [1]. The main difficulty stems from the apparent disagreement between the insignificant changes of structure and the concomitant huge variations of the dynamics across the glass transition. Whether this transition is at its core a thermodynamic phenomenon with a hidden phase transition below the experimentally observed at the glass transition temperature, $T_g$, or it is a dynamic one related to the kinetic arrest of atoms/molecules at $T_g$ is still an unsolved question[2]. A methodical experimental study of the glass transition is also challenging, due to the limited time scale range we can experimentally access.

Glasses can be seen as a mosaic of regions each one characterized by a different relaxation time, resulting from the original distribution of relaxation times in the liquid state[3]. Since glasses are out-of-equilibrium systems, they will evolve towards equilibrium, and therefore towards a new distribution of relaxation times, when the glass is annealed at a certain temperature. Whether the relaxation times will become slower or faster will be determined by the annealing temperature being, respectively, below (ageing) or above (rejuvenation[4,5], anti-aging[6] or devitrification) the limiting fictive temperature of the glass, $T_f$ [7]. This temperature is directly related to the thermodynamic stability of the glass and can be calculated as the temperature at which the state variable (enthalpy or density) has the same value for the glass and for the extrapolated equilibrium liquid [8,9]. While it has been recognized that glasses do not show a unique, macroscopic $T_f$, but rather a microscopic dispersion of a mosaic of fictive temperatures[4], it is a useful parameter to globally define the stability of a glass. Each of these mosaic regions conforming the glass will relax according to its mobility and the external temperature, producing, globally, a stretching relaxation signature with an exponent beta related to the distribution of relaxation times [3]. Mean field models like the Tool-Narayanaswamy-Moynihan (TNM) are based on this approximation to simulate the evolution of glasses under a specific thermal history [10–12]. However, this model is too simple since these regions are not isolated. One also has to consider mobility transport, as predicted by the kinetic facilitation models, that pushes/induces the relaxation in adjacent zones[13,14]. Considering both the relaxation of the different regions and the mobility transport, the glass is expected to relax progressively towards the equilibrium liquid, in a cooperative and practically homogeneous way. This type of relaxation is the one expected in liquid-cooled glasses and models based on this approach fit well enough the available experimental data [15]. However, recent works based on theoretical developments and simulations introduce one more mechanism for the transition from glass to liquid at temperatures above the limiting fictive temperature of the glass. Wolynes et al., based on the combination of Random First Order Theory (RFOT)[16,17] and Mode Coupling Theory (MCT)[18], expose that we must consider the formation of entropy drops in the glass, small mobile glassy regions that by statistical fluctuations relax fully into the equilibrated liquid and once relaxed, propagate the relaxation into the less mobile adjacent regions via a kinetic facilitation process[4]. This relaxation would spread as a flame, accelerating the transition of the glass into the liquid. The spreading of this equilibrated liquid into the adjacent regions would be faster than the time required for each of the individual regions to relax, dominating the transition of the whole glass into the liquid. Other authors have explored this alternative view of the glass transition by performing different type of simulations. In this way, Douglass and Harrowell, use the facilitated kinetic Ising model to study the relaxation of the glass into the supercooled liquid and the existence of high mobility regions is introduced as inherent dynamic heterogeneities [19]. Gutiérrez and Garrahan use local excitations to initiate the transition in the kinetically constrained model when simulating ultrastable glasses [20]. Lulli et al. introduce equilibrated higher temperature regions in a model based on a distinguishable-particle lattice, where the authors follow the spatial profiles of particle displacement and their interactions to study the relaxation of the glass [21]. Jack and Berthier use a triangular plaquette model based on spin variables considering simple interactions to reproduce the relaxation of glasses equilibrated at different temperatures (i.e. of different stability)[22]. They find that the transition in stable glasses takes place via the nucleation and growth of equilibrated liquid drops, the same transformation mechanism proposed by Wolynes et al[4]. Moreover, using the Avrami formalism[23] they successfully reproduce the dynamics of the transition. On the contrary, they find that glasses equilibrated at higher temperatures, relax via a relaxation with a broad range of relaxation times, as expected for this type of glasses. Lastly, Fullerton and Berthier [24] used the Swap Monte Carlo approach to generate in-silico glasses of ultra-high stability. The transition of this type of glasses into the liquid state takes place via the formation of liquid patches that grow until consuming the static glass matrix, following again an Avrami-like kinetics. In all these models, the requirement to observe these differentiated liquid regions is generally a big contrast in mobility between the glass and the equilibrium liquid at that temperature. This can be an explanation as why these liquid drops have not been observed experimentally in liquid-cooled glasses up to now, although according to Guiselin et al. [25], even close to $T_g$, fast equilibrated regions would form at the tail of the distribution of relaxation times, as shown in a recent simulation study where equilibrated configurations of a supercooled liquid with $\tau_\alpha \approx 100\ s$ where produced by a Swap Monte Carlo algorithm.

This theoretical framework, with a clear phase separation between the liquid and the remaining untransformed glass, has been experimentally observed in thin films of vapor-deposited ultrastable glasses, where the free surface exhibits larger mobility, and, hence, acts as a seed plane for a liquid front propagation following the kinetic facilitation concept [26–28]. The same has been possible for the bulk glass transition in ultrastable glasses either by measuring very thick films (several micrometers [29]) or by arresting the mobility of the surface [30] and avoiding in this way the formation of a propagating front [5,31]. Ultrastable glasses are prepared by means of Physical Vapor Deposition (PVD), which allows the obtention of glasses with outstanding kinetic and thermodynamic stability (i.e. higher transformation temperatures and very low fictive temperatures, respectively) [32–35], only attainable after hundreds or thousands of years of ageing of a conventional glass [27], defined as a glass obtained by cooling the liquid at q=-10 K/min. The requirements to prepare amorphous solids with such unique properties are a sufficiently slow deposition rate and a deposition temperature high enough to allow the exploration of low energy equilibrium states [27,36,37]. Experimentally, it has been found that this temperature is around $0.85T_g$ for most organic molecules[38] for deposition rates around 0.1-0.2 nm/s, being $T_g$ the glass transition temperature of the conventional glass. For these ultrastable glasses, the ratio between the average relaxation time of the glass and that of an equilibrated liquid drop at temperatures close to $T_g$, ($\tau_{glass}/\tau_\alpha$), is expected to be very large, several orders of magnitude. It has been previously shown that $\tau_{glass}$(T) can be understood as the time required to completely transform the glass into liquid at a given temperature [39,40]. We note that at $T_f$, $\frac{\tau_{glass}}{\tau_\alpha} \approx 1$. When an ultrastable glass is heated up to temperatures far above its fictive temperature, the emergence and growth of these liquid drops is apparently much faster than the relaxation of the glass and, therefore would dominate the transition of the ultrastable glass into the liquid. This is indeed what has been observed by Rodríguez-Viejo and coworkers, where ultrastable vapor deposited TPD (N,N'-Bis(3-methylphenyl)-N,N'-diphenylbenzidine) thin film glasses capped with TCTA (Tris(4-carbazoyl-9-ylphenyl)amine) were shown to transform into the liquid by the nucleation and growth of equilibrated liquid regions when submitted to isothermal treatments above the conventional glass transition temperature[5].

However, according to models and simulations, a glass with low stability, i.e one cooled from the liquid at standard rates, could also experience the bulk transition into the equilibrated supercooled liquid by forming localized liquid droplets at spots with short relaxation times provided there is sufficient mobility contrast with the adjacent regions [4,21]. Although there is not yet experimental evidence of this behavior, the measurement strategy would be to shift the transition to the supercooled liquid to temperatures much higher than its limiting fictive temperature, so the contrast in mobility between the liquid drops and the adjacent glass could be high enough for the formation and propagation of these liquid regions to be faster than the intrinsic relaxation of the surrounding glass. This assumption was already verified for surface front transformation, where experiments by Sadtchenko et al.[41] and Rodriguez-Tinoco et al.[26] showed that, at sufficient high heating rates and, hence, by sufficiently shifting the devitrification temperature of the glass, a liquid-cooled glass would also transform via front propagation.

In this article, we demonstrate experimentally that the bulk glass transition in liquid-cooled glasses can also take place via the formation of localized liquid regions instead of a cooperative relaxation under specific experimental conditions. We also observe a transition between these two mechanisms depending on the transformation temperature and the initial stability of the glass. We study the glass transition in both vapor-deposited and liquid-cooled glasses of different stabilities by carrying isothermal treatments above $T_f$. We show how all studied glasses can rejuvenate via the formation and growth of liquid regions given enough contrast in mobility between the glass and the equilibrated liquid, which in our case, translates into performing the annealing at high temperature above the crossover between $\tau_{glass}$ and $\tau_{crossover} = (180 \pm 60)\tau_\alpha$. When the transformation takes place below this crossover it develops through gradual softening, that is via a progressive relaxation of the whole glass due to the small contrast in mobility between nearby regions.

**Results and discussion**

We prepare TPD glasses ($T_g$=333 K on heating/cooling at 10 K/min) from the liquid state and from the vapor-phase to test the presence of phase separation during the transition to the super cooled liquid (SCL) depending on the stability of the glass and on the isothermal treatment above $T_f$. The glasses we analyze are: i) vapor deposited glasses grown at two different deposition temperatures: $T_{dep}$=285 K (0.85Tg), for the ultrastable glass ($T_f$=292 K) and $T_{dep}$=330 K (0.99$T_g$), a glass that grows in equilibrium with the supercooled liquid and has $T_f$=330 K. We cap both sides of the TPD glasses with TCTA ($T_g$= 428 K) to inhibit the formation of a liquid front at surfaces/interfaces [30,31]; ii) liquid-cooled glass: we deposit in the liquid state, 5 K above $T_g$ ($T_{dep}$=338K), we then cool it down to room temperature and then we age the resulting glass at $T_{ann}$=319 K ($T_g$-14 K) for 96 h until it is completely equilibrated at that temperature ($T_f$=319K). We cap it afterwards with TCTA to avoid the formation of the liquid front on the surface during the final upscan (see figure 1). Since the sample is directly deposited in the liquid state well above $T_g$, this approach allows us to generalize the observation of phase separation during the glass to liquid transition to both vapor-deposited and liquid-cooled glasses. The samples are directly deposited onto the membrane of a nanocalorimeter which allows us to apply the thermal protocol shown in figure 1 right after growth without breaking vacuum. After deposition the glasses are taken to a certain temperature above the $T_f$ of that specific glass and remain there for different times. During this annealing treatment, the glass partially rejuvenates. We then cool down the sample at approximately -500 K/s and perform a subsequent fast heating ($\beta \approx 3.5 \times 10^4$ K/s) scan with the nanocalorimeter. The heat capacity traces of the final fast upscan are therefore representative of the thermal history of the glass after the isothermal treatment. The fast cooling/fast heating attained with our custom-made nanocalorimetric chips is a key point to resolve the heat capacity overshoots of glassy zones with different stabilities. That is, if some liquid regions are formed during the previous isotherm, they will become glass regions of very low stability because of the fast cooling rate employed[5]. In this case, two glass transition signatures will be observed during heating, one at lower temperature for the very low stability glass and one at higher temperature, corresponding to the untransformed (or partially relaxed) glass during the isotherm.

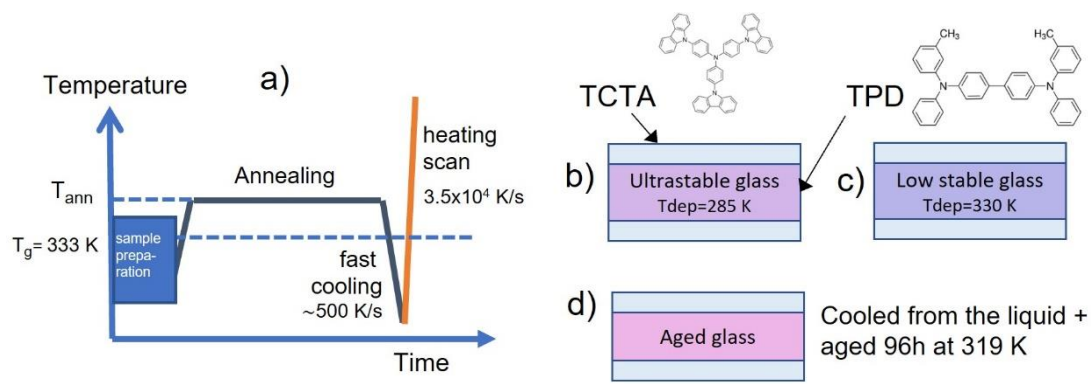

Figure 1. a) Thermal treatment performed on the different samples. After sample preparation, samples are annealed at $T_{ann}$ for different times. After cooling them down at ~-500 K/s, heat capacity data is recorded at a heating rate of $3.5 \times 10^4$ K/s. Different samples under study: b) ultrastable vapor deposited glass grown at $T_{dep}$=285 K (0.85$T_g$); c) low stability vapor deposited

glass grown at $T_{dep}$=330 K (0.99$T_g$); d) glass obtained after depositing 5 K above $T_g$, cooled down to 319 K and aged there until complete stabilization (96h).

Figure 2 shows the specific heat curves for a TPD glass deposited at $T_{dep}$=330 K ($T_f$=330 K) after isothermal treatments at 341K (Tg+8K) and 347 K (Tg+14K) for different times. The as-deposited glass, grown in equilibrium with the supercooled liquid at $T_{dep}$, is comparable to a glass obtained by cooling the liquid at a cooling rate around 1 K/min. The transformation of this glass shows remarkable differences depending on the annealing temperature. Annealing at $T_{ann}$=341 K results in a shift of the glass transition peak towards lower temperatures as time increases, indicating a progressive relaxation of the glass towards the equilibrium liquid at that specific annealing temperature. On the contrary, at $T_{ann}$=347 K we see both the relaxation of the as-deposited glass, by the shift of the original peak to lower temperatures, and the formation of distinct liquid regions in the glass that are manifested through the apparition of a second peak at lower temperature[5] identified by a green arrow in Figure 2a and a cartoon representing the formation and growth of the liquid drops. We would like to emphasize the relevance of what is shown in figure 2a. The same glass can experience different transformation mechanisms during the glass transition depending on the temperature at which the transition takes place. Remarkably, even that both temperatures differ just by 6 K, the contrast in mobility between liquid and glass is apparently different enough at these two temperatures to show distinct outputs when the transformation is partially fulfilled.

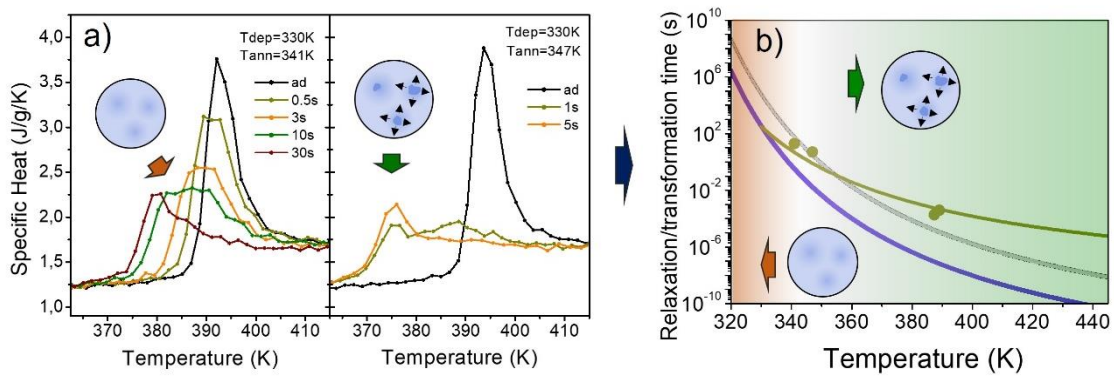

Figure 2. a) Specific heat traces obtained at a scanning rate of $3\times10^4$ K/s after different annealing treatments of samples deposited at 0.99$T_g$. The annealing temperatures and times are indicated in the legend. The endothermic peak that appears at lower temperatures is an indication of the formation of liquid patches in the glass during the annealing treatment. b) curves corresponding to the alpha relaxation time of the liquid (blue) [42], the relaxation time of the glass (green), which

*depends on the limiting fictive temperature of the glass ($T_f$=330 K) and has been calculated according to Rodriguez-Tinoco et al[39] (green), the line corresponding to the crossover relaxation time calculated as $\tau_{glass} = (180 \pm 60)\tau_\alpha$ (grey). In orange, the temperature regions for which a cooperative relaxation mechanism is expected for the glass transition. In green, the temperature region where we expect to find the nucleation and growth of liquid regions during the glass transition. The cartoons represent the two transformation mechanisms.*

We propose the appearance of equilibrated regions can be rationalized from the ratio between the relaxation time of the glass and the alpha relaxation time of the liquid ($\tau_{glass}/\tau_\alpha$) at the temperature of the isothermal treatment. If this ratio is large, the nucleation and growth of liquid drops dominates the transition[5] since on average the relaxation of the glass is slower than the rate at which the liquid emerges and consumes the glass. On the contrary, small ratios are an indication that a cooperative heterogenous dynamics across the sample is fast enough to be the main active mechanism during the transition. To further analyze this assumption and provide a numeric estimation, we represent in figure 2b the experimental transformation times of the glass deposited at $T_{dep}$=0.99$T_g$ as a function of temperature (green circles). A detailed explanation on how to calculate these times can be found in the methods section. In the same graph, we have also plotted the alpha relaxation time (blue line) and the relaxation/transformation time of the glass (green line), calculated as explained elsewhere[39] and briefly addressed at the methods section. In this particular case we see a change of mechanism in a very small range of temperatures, so we assume that the crossover temperature between the two regimes, gradual softening (orange colored) and formation of distinctive liquid regions (green colored), should be at a temperature between these two. For the annealing at 341 K, $\tau_{glass}/\tau_\alpha \approx 60$, while at $T_{ann}$=347 K, $\tau_{glass}/\tau_\alpha \approx 300$. As a rough estimation, we take the average between these two values as the crossover ratio. Figure 2b shows as a grey line this crossover relaxation time, $\tau_{glass,crossover} = (180 \pm 60)\tau_\alpha$. We use this value as a predictor to identify the temperature required to observe the formation of isolated liquid regions during the glass transition for a specific glass irrespective of the initial state of the glass. The transformation time at 341 K falls to the left of the crossover line, while the one at 347 K will be at the right side, showing at each temperature different transformation mechanisms.

The temperature of the crossover is somewhat ill-defined because of the difficulty to establish in this region a clear difference between both mechanisms in the heat capacity traces and therefore we represent it by a broad white-graded area (see figure 2b). The cartoons (together with background color) in figure 2b clearly illustrate the impact of the ratio $\frac{\tau_{glass}}{\tau_\alpha}$ on the transformation, and schematically represent the transformation at both sides, showing regions of liquids drops at the right (darker blue) within a glassy matrix (softer blue), while in the left region the transformation is spatially less resolved.

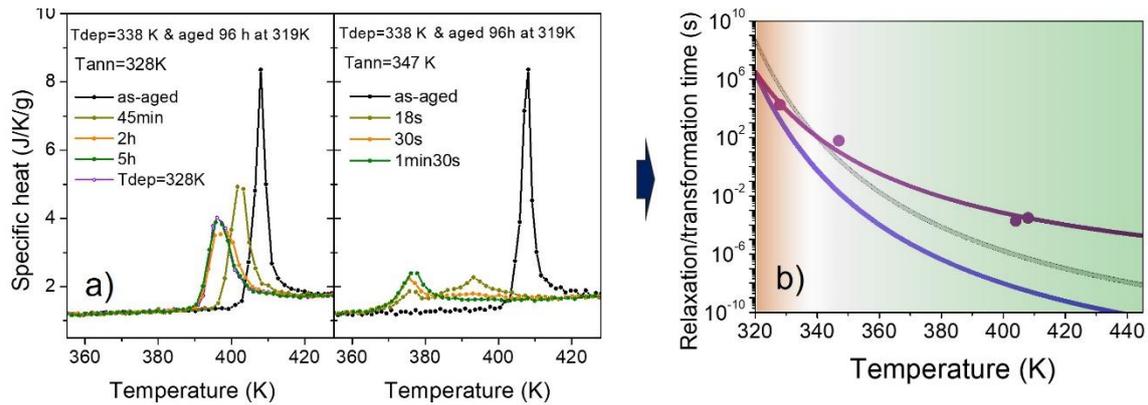

Figure 3. *a) Specific heat traces after different annealing treatments of samples deposited at 1.05Tg, fast cooled and aged at 319K for 4 days. The annealing temperatures and times are indicated in the legend. Left pannel: The overlap between the curve corresponding to a sample deposited at 328K and the one annealed for 5h at 328 K indicates that after this time, the glass has fully transformed and reached equilibrium. b) curves corresponding to the alpha relaxation time of the liquid (blue)* [42]*, the relaxation time of the glass (purple), and the line corresponding to the crossover relaxation time calculated as* $\tau_{glass,crossover} = (180 \pm 60)\tau_\alpha$ *(grey). Violet circles correspond to experimental data.*

A similar behavior has been observed in a glass cooled from the liquid. Figure 3a shows the specific heat traces of a glass deposited 5 K above $T_g$ (338 K), cooled down and aged at 319 K for 96 h. This treatment led to full equilibration of the glass at this temperature, i.e. $T_f$=319 K, before performing the isothermal treatment at 347 K (Tg+14 K) and at Tg-5 K (328 K). At both temperatures the glass is expected to evolve towards the supercooled liquid, which means towards faster relaxation times [7]. As can be seen in the calorimetric traces, during the annealing at 347 K, distinguishable liquid regions are formed as part of the transition of the glass into the supercooled liquid. On the contrary, when the annealing is performed at 328 K, there is a continuous shift of the glass transition overshoot, with no distinguishable fast-cooled glassy regions, and the transition takes place exclusively via a cooperative relaxation process (gradual softening). In figure 3b we represent the relaxation times of the liquid and the glass (for a glass with $T_f$=319K), the transformation times and the crossover relaxation time as in figure 2b. As can be seen in the figure, the transformation times at 328 K (cooperative heterogenous dynamics on the whole sample) and 347 K (localized heterogenous dynamics + facilitation) fall respectively at left and right of the crossover line ($\tau_{glass,crossover} = (180 \pm 60)\tau_\alpha$), consistent with the transition mechanism observed during the annealing treatments (figure 3a) and providing solidity to the assumption that this crossover curve is independent of the characteristics of the glass. According to our estimation, for this particular glass, since its stability is high ($T_f$=319 K), the annealing temperatures required to exclusively see the cooperative relaxation would be below 338 K. In fact, the higher the stability of the glass, the lower the crossover temperature. This is even clearer in the case of the ultrastable glass. Figure 4a shows the calorimetric traces for a sample vapor-deposited at $T_{dep}$=285 K (0.85 Tg), which is the temperature at which the maximum kinetic and thermodynamic stability is attained [31,43], after annealing at 347 K (Tg+14 K) for different times. At this temperature the ratio $\tau_{glass}/\tau_\alpha \approx 2 \cdot 10^6$ and the transformation is clearly dominated by the nucleation and growth of liquid patches in the glass [5]. Looking at a similar scheme than in figures 2b and 3b, one can see in the inset of figure 4b that we would need to attain T<300 K (33 K below $T_g$) with unreachable transformation times above $10^{20}\tau_\alpha$ to transform the glass by the cooperative gradual softening.

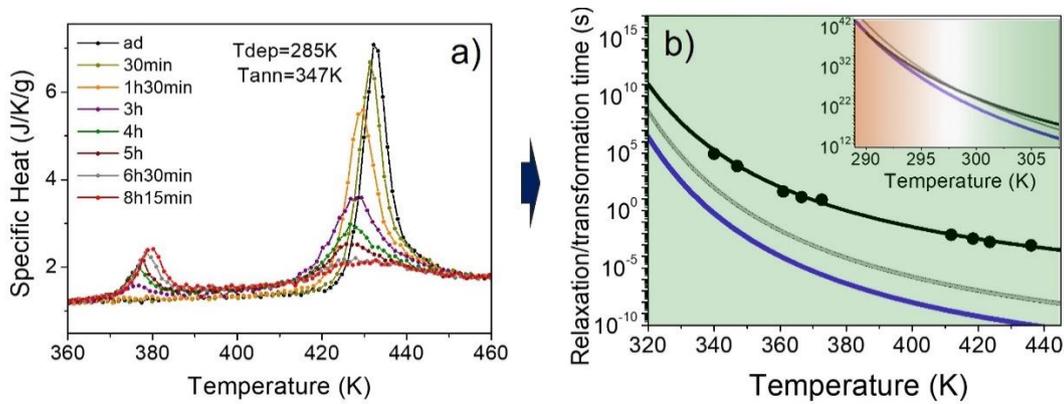

*Figure 4. a) Specific heat traces after different annealing treatments of samples deposited at 0.85Tg (285 K) and annealed at 347 K. The annealing times are indicated in the legend. b) curves corresponding to the alpha relaxation time of the liquid (blue) [42], the relaxation time of the glass (black), and the line corresponding to the crossover relaxation time calculated as $\tau_{glass} = (180 \pm 60)\tau_\alpha$ (grey). The inset shows the same curves but in a lower temperature range, so the crossover point can be depicted.*

The crossover line could be rationalized by considering the intrinsic dynamical heterogeneity of the supercooled liquid state. Even though the characteristic time of the relaxation of a liquid is monitored via a single value of structural relaxation for each temperature, the true relaxation proceeds via non-stretched exponential decay of dynamic correlations in spontaneous density fluctuations. The relaxation dynamics of the SCL is typically assessed by dielectric spectroscopy, where this relaxation is manifested as a peak in the complex part of the permittivity of the sample (dielectric loss). While the maximum of this peak is regarded as the (main) relaxation time of the system, its width is strongly influenced by the dynamical heterogeneities in the SCL. As a consequence, even after a time $\tau_\alpha$, parts of the SCL are still not relaxed, corresponding to the low frequency side of the dielectric loss, which for most glass-formers, but in particular for TPD, spreads along two orders of magnitude above the maximum of the peak (alpha relaxation value)[44]. Interestingly, this value is consistent with the observation of the crossover time of $(180 \pm 60)\tau_\alpha$. Under this view, we could interpret that if the relaxation time of the glass is inside the distribution of relaxation times of the liquid, the whole system relaxes as a SCL would do, i.e., via heterogeneous relaxation mechanism with no nucleation of liquid clusters.

The relaxation time of the glass at the crossover temperature may have been the major handicap for observing experimentally these liquid regions during the glass transition. The lower the stability of the glass, the higher the crossover temperature and, therefore, the faster the transformation time of the glass. In the case of the glass deposited at 0.99$T_g$, for instance, the crossover occurs for transformation times of the order of a few seconds. For a conventional glass cooled at -10 K/min with $\tau_\alpha \approx 100\,s$ at $T_g$ 'nucleation and growth' behavior would be identified only at temperatures above 359 K, with transformation times around several ms. Since conventional calorimetry is unable to work in these short time scales, most of the experiments up to now were limited to temperatures near $T_g$ (at the left side of the crossover) and therefore, only the cooperative relaxation mechanism had been observed so far. Now with the general availability of fast scanning calorimetry short time scales are within reach and we challenge other experimental groups to carry out measurements far from equilibrium to test the occurrence of similar behavior in other glassy systems.

Our work provides conclusive experimental evidence to test theories of the glass transition and is compatible with models that include heterogeneous dynamics and dynamic facilitation and predict the formation of distinct liquid regions when there exists a large contrast between the mobility of the equilibrated liquid and the glass. We have also shown that the ratio of $\tau_{glass}/\tau_\alpha$ at a given temperature appears to be a good indicator to predict the mechanism of the transformation. Importantly, the existence of the two regimes identified in this work is independent of the experimental procedure or protocol to produce the glass but their exploration may require the use of ultrafast experimental techniques to identify the liquid regions.

**Acknowledgements**

JRV and MGS acknowledge Grant MAT2016-79759-R funded by MCIN/AEI/ 10.13039/501100011033 and "ERDF A way of making Europe", and Grant PID2020-117409RB-I00 funded by MCIN/AEI/ 10.13039/501100011033. CRT is a Serra Húnter Fellow. The ICN2 was funded by the CERCA programme / Generalitat de Catalunya. The ICN2 was supported by the Severo Ochoa Centres of Excellence Programme, funded by the Spanish Research Agency (AEI, Grant no. SEV-2017-0706). All the authors acknowledge L. Abad and IMB-CNM for the fabrication of the nanocalorimeters.

**Methods:**

TCTA, with a Tg 91 K above the one of TPD (Tg=333 K), is a good candidate to arrest the mobility of the TPD surfaces, as has been shown previously [31]. So, in order to access the bulk transformation of the TPD glass, a 13 nm TCTA layer has been deposited at the bottom and the top of the TPD layers, with thicknesses around 65 nm. The TPD/TCTA sandwich has been deposited on the active zone of the membrane based calorimetric chips by means of physical vapor deposition. The deposition chamber has a N2 trap to improve the vacuum and to act as a heat sink for the fast cooling of the samples. The base pressure is of the order of $10^{-8}$ mbar. Two evaporators and two shutters allow the sequential evaporation of the two materials without breaking the vacuum. The thickness is controlled by a previously calibrated quartz crystal monitor. The temperature control during deposition and annealing treatment is performed with the calorimetric chips, by providing a specific intensity, which will heat the Pt circuit on the membrane. Heat scans are performed by introducing a short pulse of intensity in the chips, which will induce a constant heating ramp. In this work, pulses of 35 mA have been used, which result in heating rates of around $3.5 \cdot 10^4$ K/s. Data of the resistivity of the chip as a function of time can be processed to obtain the heat capacity as function of temperature. More information about the technique can be found in [45,46]. Specific heat data for TPD has been obtained first subtracting the heat capacity contribution of glassy TCTA to the heat capacity curves and dividing them by the TPD mass.

In order to calculate the relaxation/transformation time of the glass we use two different approaches as explained elsewhere [39]. In the first approach, we employ the expression $\tau_1 \beta_1 = \tau_2 \beta_2$ [47] to obtain the value of the glass relaxation time from the heating rate of the experiment, assigning this value to the onset temperature of the glass transition. As reference we use a value for the relaxation time of the glass of 100 s for a heating rate of 10 K/min [48,49]. In the second approach, we calculate the transformation time from the width of the transformation peak, and the corresponding value of the heating rate, assigning it to the temperature at the maximum of the transformation peak: $t_{trans}(T_{max}) = \Delta T/\beta$, where $\Delta T$ is the width of the transformation peak and $\beta$ the value of the heating rate during the transformation. Both approaches yield comparable results. Previous works have already considered this equivalence [39,50]. The transformation/relaxation time curve as function of temperature corresponds to the equation $\tau_g = \tau_{g0} \exp\left[\frac{\xi(T_f') T_0}{(T - T_0)}\right]$, which is a generalization of the well-known Vogel-Fulcher-Tamman, VFT, equation [51], aimed at describing the dynamics of supercooled liquids and glasses with different thermal stability. In this equation all the parameters have an analogous meaning as in VFT equation. In this case, however, D has been substituted by a linear function of the limiting fictive temperature of the glass, $\xi(T_f') = AT_f' + B$. In a supercooled liquid the fictive temperature $T_f$ = T at all temperatures, from the definition of $T_f$. More information of the validity of this equation can be found in [39].